\documentstyle[12pt,epsf]{article}

\begin{document}

\renewcommand{\theequation}{\thesection.\arabic{equation}}
\newcommand{\reseteqnum}{\setcounter{equation}{0}}

\title{
\hfill
\parbox{4cm}{\normalsize UT-950\\
KEK Preprint 2001-53\\
hep-th/0107217}\\
\vspace{1cm}
Stability of Quiver Representations\\
and Topology Change
\vspace{1cm}}
\author{Tomomi Muto\thanks{e-mail address:\tt 
muto@hep-th.phys.s.u-tokyo.ac.jp}$\;$
and Taro Tani\thanks{e-mail address:\tt tanitaro@post.kek.jp}
\vspace{0.5cm}\\
{\normalsize\em $^*$Department of Physics, 
University of Tokyo}\\
{\normalsize\em Hongo 7-3-1, Tokyo 113-0033, Japan}
\vspace{0.5cm}\\
{\normalsize\em $^\dagger$Institute of Particle and 
Nuclear Studies}\\
{\normalsize\em High Energy Accelerator Research 
Organization(KEK)}\\
{\normalsize\em Oho 1-1, Tsukuba, Ibaraki 305-0801, Japan}}
\date{\normalsize}
\maketitle

\begin{abstract}
\normalsize
We study phase structure of the moduli space of a D0-brane 
on the orbifold ${\bf C}^3/{\bf Z}_2 \times {\bf Z}_2$ 
based on stability of quiver representations.
It is known from an analysis using toric geometry that this 
model has multiple phases connected by flop transitions.
By comparing the results of the two methods, we obtain 
a correspondence between quiver representations and 
geometry of toric resolutions of the orbifold.
It is shown that a redundancy of coordinates arising 
in the toric description of the D-brane moduli space, 
which is a key ingredient of disappearance of non-geometric 
phases, is understood from the monodromy around the orbifold 
point.
We also discuss why only geometric phases appear from 
the viewpoint of stability of D0-branes.

\end{abstract}

\newpage
\section{Introduction}

D-branes allow us to study the structure of geometry at 
sub-stringy scales.
From this perspective, space-time is a derived concept 
appearing as a vacuum moduli space of the D-brane worldvolume 
gauge theory.
A number of works have been devoted to investigating geometry 
as seen by D-branes and comparing it with standard classical 
geometry or geometry probed by fundamental strings.

In \cite{DM}, it was pointed out that D-branes on orbifolds 
${\bf C}^2/\Gamma$ are described by quiver gauge theories, 
where a quiver is a diagram representing algebraic structure 
of the group $\Gamma$.
It implies that D-branes give a physical explanation of 
the McKay correspondence: a relation between geometry of 
resolutions of an orbifold ${\bf C}^n/\Gamma$ and the 
representation theory of $\Gamma$.
 (For mathematics on the McKay correspondence, 
see \cite{McKay}-\cite{BKR}.)
The purpose of this paper is to study a relation between 
geometry of an orbifold ${\bf C}^n/\Gamma$ probed by 
a D0-brane and the representation theory of $\Gamma$ 
based on the concept of stability of the D0-brane.

Stability has been recently brought to attention as 
it plays an important role in studying D-branes 
on Calabi-Yau manifolds~\cite{DFR1}-\cite{Fiol}.
The relevant notion of stability near the orbifold point 
in the K${\rm {\ddot a}}$hler moduli space is 
$\theta$-stability of representations of quivers~\cite{King}.
It enables us to construct the vacuum moduli space of the 
quiver gauge theories without solving all equations defining 
the vacua and to study the marginal stability loci, 
where D-brane spectrum jumps.

We would like to compare the results obtained from 
stability of quiver representations corresponding to 
a D0-brane with geometric structure of three-dimensional 
orbifolds ${\bf C}^3/\Gamma$ with $\Gamma$ an abelian subgroup 
of $SU(3)$.
The latter is investigated following the procedure given 
in \cite{DGM}.
In this formulation, the vacuum moduli space is obtained 
by solving both F-flatness and D-flatness conditions, 
the equations defining the vacua, with the aid of 
toric geometry.
The D-flatness conditions contain Fayet-Iliopoulos terms 
originating from twisted sectors of closed strings, 
which parameterize the K${\rm {\ddot a}}$hler moduli space 
of the orbifold.
The core of the method in \cite{DGM} is to convert 
F-flatness conditions into D-flatness conditions of 
an auxiliary gauge theory with a large number of 
chiral multiplets.
Scalar components of the chiral multiplets become homogeneous 
coordinates describing the vacuum moduli space.
After reducing redundant coordinates by using D-flatness 
conditions and gauge symmetry, we obtain the vacuum moduli 
space of the quiver gauge theory.

Various models were investigated following 
this procedure~\cite{Muto}-\cite{Sarkar}.
The most striking feature common to these analyses is that 
non-geometric phases are projected out.
This is in contrast to the analyses based on 
fundamental strings~\cite{AGM1,Witten}, 
in which non-geometric phases are realized as 
abstract conformal field theories.
By inspecting the process of the calculation of the 
D-brane moduli spaces, we can see that the key to the 
disappearance of non-geometric phases is the redundancy of 
coordinates mentioned above.
So far, however, physical origin of the redundancy has not 
been clarified; it just arises as a result of a combinatorial 
algorithm converting F-flatness conditions to D-flatness 
conditions.
To understand this point is one motivation of this work.

The orbifold we study in this paper is 
${\bf C}^3/{\bf Z}_2 \times {\bf Z}_2$.
A reason to study this orbifold is that it has a rather 
rich structure in spite of its simplicity; 
it contains multiple phases that are related by 
topology changing processes called flops~\cite{Greene,MR} 
as we review in section 2. 
We also see that non-geometric phases disappear due to 
the redundancy of homogeneous coordinates.
In section 3 we re-examine the moduli space of D0-branes 
based on $\theta$-stability of quiver representations.
By comparing the results with those obtained in section 2, 
we obtain a correspondence between homology cycles of a 
resolution of the orbifold and quiver representations.
Furthermore, we find that the redundancy of the homogeneous 
coordinates stems from monodromy around the orbifold point 
in the K${\rm {\ddot a}}$hler moduli space.
Finally, we discuss the disappearance of non-geometric phases 
from the viewpoint of stability of D0-branes.

\section{D-branes on ${\bf C}^3/{\bf Z}^2 \times {\bf Z}^2$: 
an analysis based on toric geometry}
\reseteqnum

In this section, we first review the method given in 
\cite{DGM} to study D0-branes on orbifolds ${\bf C}^3/\Gamma$ 
with $\Gamma \in SU(3)$.
Second, we present the geometric structure of the orbifold 
${\bf C}^3/{\bf Z}^2 \times {\bf Z}^2$ obtained by this 
method~\cite{Greene,MR}.

D-branes on orbifolds ${\bf C}^3/\Gamma$ are described by 
quiver gauge theories.
A quiver is a graph consisting of a set of nodes $v_i \in V$ 
and a set of arrows $a_{ij} \in A$ starting from the node $v_i$ 
and ending at the node $v_j$.
We will denote the number of the nodes as $N$.
Given a quiver, one obtains a gauge theory by considering 
a representation of the quiver.
A representation $R$ of a quiver is a collection of finite 
dimensional vector spaces $V_i$, one for each node $v_i \in V$, 
and a collection of linear maps $X_{ij}:V_i \rightarrow V_j$, 
one for each arrow $a_{ij} \in A$.
The vector $(n_0,n_1,...,n_{N-1})$, where $n_i$ is the 
dimension of the vector space $V_i$, is referred to as the 
dimension vector of the representation $R$.

The gauge theory corresponding to a representation $R$ is 
an ${\cal N}=1$ supersymmetric gauge theory with a gauge 
symmetry $G=\Pi_{v_i \in V} U(n_i)/U(1)$: a node $v_i$ 
represents a factor of $U(n_i)$ in the gauge group $G$, 
and an arrow $a_{ij}$ represents a chiral multiplet 
transforming as $(n_i,{\bar n_j})$ under $U(n_i) \times U(n_j)$.
Note that the factor $U(1)$ in the gauge group is the diagonal 
subgroup of $\Pi_{v_i \in V} U(n_i)$ which acts trivially.

A quiver relevant to the discussion of D-branes on 
${\bf C}^3/\Gamma$ is the McKay quiver associated to $\Gamma$.
Nodes in the McKay quiver correspond to irreducible 
representations of $\Gamma$, and arrows encode information 
on tensor products of a three-dimensional representation 
defining the action of $\Gamma$ on ${\bf C}^3$ and irreducible 
representations of $\Gamma$.
In this paper, we restrict $\Gamma$ to be an abelian subgroup 
of $SU(3)$, whose irreducible representations are 
one-dimensional.
In that case, the number $N$ of nodes in the McKay quiver 
coincides with the order of $\Gamma$.
Thus a D0-brane on an orbifold ${\bf C}^3/\Gamma$, 
which we would like to discuss, 
corresponds to a quiver representations with a dimension 
vector $(1,1,...,1)$.

We are concerned with the vacuum moduli space $\cal M$ of 
the quiver gauge theory.
The vacua are parameterized by values for scalars in 
the chiral multiplets, modulo gauge equivalence, solving 
F-flatness and D-flatness conditions.
In \cite{DGM}, it was found to be convenient to consider 
F-flatness conditions as if they were D-flatness conditions 
in an auxiliary gauge theory.
The vacuum moduli space of this auxiliary gauge theory is a 
$(N+2)$-dimensional space of the following form,
\begin{equation}
\{(p_1,...p_c) \in {\bf C}^c|(c-N-2) \; \mbox{D-flatness 
conditions}\} /U(1)^{c-N-2},
\label{eq:FD}
\end{equation}
where $c$ is the number of scalars in the chiral multiplets of 
the auxiliary gauge theory.
The data necessary to construct this space are obtained through 
a combinatorial algorithm based on toric geometry.
This process is burdensome in general, and hence an analytic 
expression of $c$ is not known for ${\bf C}^3/{\bf Z}_n$ nor 
${\bf C}^3/{\bf Z}_n \times {\bf Z}_m$.

To obtain the vacuum moduli space ${\cal M}$ of the quiver 
gauge theory, one must further impose D-flatness conditions
\begin{equation}
D_i=\sum_j (X_{ji}^\dagger X_{ji}-X_{ij}X_{ij}^\dagger )=
\theta_i,
\label{eq:D-flatness}
\end{equation}
that exist in the quiver gauge theory from the beginning.
Note that although the index $i$ takes values from 0 to $N-1$, 
the number of independent conditions is $N-1$.
Real parameters $\theta_i$ come from twisted sectors of NS-NS 
fields, and they are related to physical Fayet-Iliopoulos 
parameters $\zeta_i$ through the following relation:
\begin{equation}
\theta_i=\zeta_i-\frac{\sum_k \zeta_k n_k}{\sum_l n_l}.
\end{equation}
This relation comes from the requirement of quasi-supersymmetry 
of the vacuum~\cite{DFR1}.
In the case of $(n_0,n_1,...,n_{N-1})=(1,1,...,1)$, $\theta_i$ 
coincide with the physical Fayet-Iliopoulos parameters 
$\zeta_i$.

The total number of D-flatness conditions to be imposed on the 
auxiliary gauge theory is $(c-N-2)+(N-1)=c-3$, and gauge 
symmetry is $U(1)^{c-N-2} \times U(1)^{N-1}=U(1)^{c-3}$.
Thus the moduli space $\cal M$ takes the form
\begin{equation}
{\cal M}=\{(p_1,...,p_c) \in {\bf C}^c|(c-3) \; 
\mbox{D-flatness conditions}\}/U(1)^{c-3}.
\label{eq:moduli}
\end{equation}
Note that the $N-1$ D-flatness conditions in equation 
($\ref{eq:D-flatness}$) have Fayet-Iliopoulos parameters 
$\theta_i$, while $c-N-2$ D-flatness conditions coming from 
F-flatness conditions do not have such parameters.

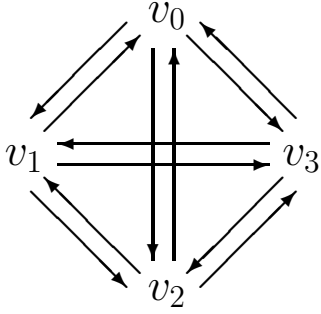
\begin{figure}[htbp]
\begin{center}
\begin{picture}(120,120)
\put(54,110){\Large $v_0$}
\put(0,56){\Large $v_1$}
\put(54,5){\Large $v_2$}
\put(105,56){\Large $v_3$}
\thicklines
\put(20,56){\vector(1,0){80}}
\put(100,64){\vector(-1,0){80}}
\put(64,20){\vector(0,1){80}}
\put(56,100){\vector(0,-1){80}}
\put(10,46){\vector(1,-1){36}}
\put(110,74){\vector(-1,1){36}}
\put(74,10){\vector(1,1){36}}
\put(46,110){\vector(-1,-1){36}}
\put(15,69){\vector(1,1){36}}
\put(105,51){\vector(-1,-1){36}}
\put(51,15){\vector(-1,1){36}}
\put(69,105){\vector(1,-1){36}}
\end{picture}
\caption{The quiver diagram for ${\bf Z}_2 \times {\bf Z}_2$.}
\label{fig:quiver}
\end{center}
\end{figure}

Now we investigate the case ${\bf C}^3/{\bf Z}_2 \times 
{\bf Z}_2$ in detail~\cite{Greene,MR}.
Non-trivial elements $g_1,g_2$ $g_3$ of $\Gamma={\bf Z}_2 
\times {\bf Z}_2$ act on the coordinates $(z_1,z_2,z_3)$ of 
${\bf C}^3$ as follows:
\begin{eqnarray}
g_1:(z_1,z_2,z_3) \rightarrow (z_1,-z_2,-z_3),\nonumber\\
g_2:(z_1,z_2,z_3) \rightarrow (-z_1,z_2,-z_3),\\
g_3:(z_1,z_2,z_3) \rightarrow (-z_1,-z_2,z_3).\nonumber
\end{eqnarray}
Here $g_1^2=g_2^2=g_3^2=1,~g_1 g_2=g_3$.
Since ${\bf Z}_2 \times {\bf Z}_2$ is an abelian group of order 
four, there are four irreducible representations with dimension 
one.
The quiver diagram is depicted in Figure \ref{fig:quiver}.

Since the quiver has four nodes ($N=4$), the associated quiver 
gauge theory has a gauge symmetry $U(1)^3$ and three 
Fayet-Iliopoulos parameters $\theta_i$.
The vacuum moduli space of the quiver gauge theory is computed 
following the procedure stated above.
In this case, the combinatorial algorithm gives an auxiliary 
gauge theory with $c=9$.
Therefore the vacuum moduli space ${\cal M}$ is represented as,
\begin{equation}
{\cal M}=\{(p_1,...p_9) \in {\bf C}^9|6 \; 
\mbox{D-flatness conditions}\}/U(1)^6,
\end{equation}
where six D-flatness conditions are
\begin{eqnarray}
-|p_4|^2+|p_7|^2 &=& 2\tilde \theta_1, \label{eq:D-flat1}\\
-|p_5|^2+|p_8|^2 &=& 2\tilde \theta_2, \label{eq:D-flat2}\\
-|p_6|^2+|p_9|^2 &=& 2\tilde \theta_3, \label{eq:D-flat3}\\
|p_1|^2+|p_4|^2-|p_5|^2-|p_6|^2 &=&
-\tilde \theta_1+\tilde \theta_2+\tilde \theta_3, 
\label{eq:D-flat4}\\
|p_2|^2-|p_4|^2+|p_5|^2-|p_6|^2 &=&
\tilde \theta_1-\tilde \theta_2+\tilde \theta_3, 
\label{eq:D-flat5}\\
|p_3|^2-|p_4|^2-|p_5|^2+|p_6|^2 &=&
\tilde \theta_1+\tilde \theta_2-\tilde \theta_3. 
\label{eq:D-flat6}
\end{eqnarray}
Here we have used parameters $\tilde \theta_1 = 
(\theta_2+\theta_3)/2,~\tilde \theta_2 = (\theta_3+\theta_1)/2,~
\tilde \theta_3 = (\theta_1+\theta_2)/2$.

\begin{figure}[htdp]
\begin{center}
\leavevmode
\epsfysize=45mm
\epsfbox{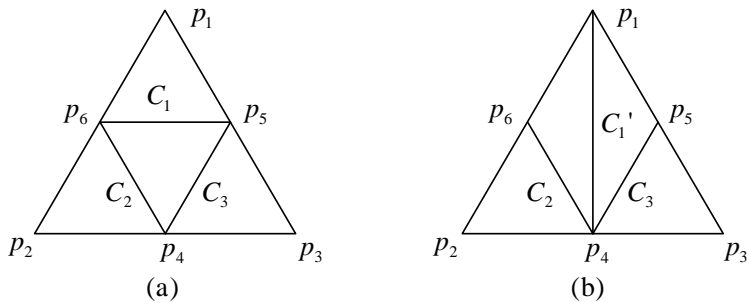}
\caption{Toric diagrams representing the topology of the 
resolved orbifold (a) in the region $\theta_1>0,~\theta_2>0,~
\theta_3>0$, and (b) in the region $\theta_1<0,~
\tilde \theta_2=(\theta_1+\theta_3)/2>0,~
\tilde \theta_3=(\theta_1+\theta_2)/2>0$.}
\label{fig:toric}
\end{center}
\end{figure}

We first study the geometric structure of $\cal M$ in the region
$\tilde \theta_1>0$, $\tilde \theta_2>0$, $\tilde \theta_3>0$.
For $\tilde \theta_1>0$,
$p_7$ is nonzero due to the equation (\ref{eq:D-flat1}).
Combining with a U(1) gauge symmetry corresponding to this 
D-flatness condition, we can eliminate $p_7$.
Similarly, in the region $\tilde \theta_2>0$ 
($\tilde \theta_3>0$), we can eliminate $p_8$ 
(respectively $p_9$).
Thus in the region $\tilde \theta_1>0$,
$\tilde \theta_2>0$ and $\tilde \theta_3>0$, the six D-flatness 
conditions are reduced to the last three equations 
(\ref{eq:D-flat4}), (\ref{eq:D-flat5}), (\ref{eq:D-flat6}).
They describe a resolution of ${\bf C}^3/{\bf Z}_2 \times 
{\bf Z}_2$.
However, the topology of the resolution is not fixed uniquely 
since the inequalities $\tilde \theta_1>0$, $\tilde \theta_2>0$,
$\tilde \theta_3>0$ do not fix the sign of the right-hand side 
of (\ref{eq:D-flat4}), (\ref{eq:D-flat5}), (\ref{eq:D-flat6}); 
flop transition occurs as the sign of the right-hand side 
changes.
For example, in the region 
$\theta_1=-\tilde \theta_1+\tilde \theta_2+\tilde \theta_3>0,~
\theta_2=\tilde \theta_1-\tilde \theta_2+\tilde \theta_3>0,~
\theta_3=\tilde \theta_1+\tilde \theta_2-\tilde \theta_3>0$, 
the resolved orbifold has a topology represented by the toric 
diagram drawn in Figure \ref{fig:toric}(a).
Each line in the diagram represents homology 2-cycle $C_i$ in 
the resolution of ${\bf C}^3/{\bf Z}_2 \times {\bf Z}_2$, 
whose volume is parameterized by $\theta_i$.
If we change the sign of $\theta_1$, $C_1$ is replaced by 
$C'_1$ as depicted in Figure \ref{fig:toric}(b).
The volume of $C'_1$ is parameterized by $-\theta_1$, 
which implies that the homology class $[C'_1]$ is equal to 
$-[C_1]$ as noted in \cite{AGM2}.

Before we come to the analysis outside of $\tilde \theta_1>0,~
\tilde \theta_2>0,~\tilde \theta_3>0$, we comment on the phase 
structure obtained from just the three D-flatness conditions 
(\ref{eq:D-flat4}), (\ref{eq:D-flat5}), (\ref{eq:D-flat6}).
These are the D-flatness conditions that appear in the analysis 
of the orbifold ${\bf C}^3/{\bf Z}_2 \times {\bf Z}_2$ based on 
fundamental strings~\cite{Witten}.
In the region $\tilde \theta_1>0,~\tilde \theta_2>0,~
\tilde \theta_3>0$, one obtains the same resolutions of the 
orbifold as described above.
In another region, however, one obtains a singular space.
Such regions are called non-geometric phases realized as 
abstract conformal field theories.

\begin{figure}[htdp]
\begin{center}
\leavevmode
\epsfysize=70mm
\epsfbox{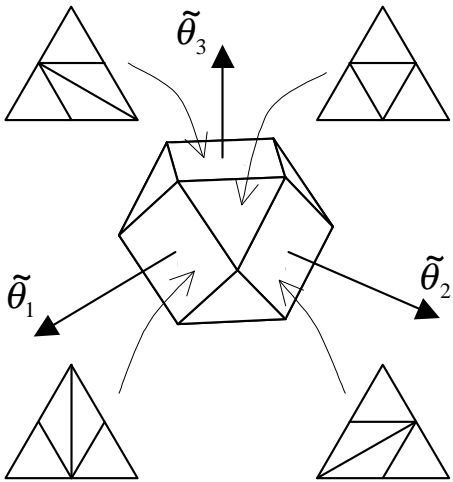}
\caption{The phase structure of the orbifold 
${\bf C}^3/{\bf Z}_2 \times {\bf Z}_2$ probed by a D0-brane 
near the orbifold point of the K${\rm {\ddot a}}$hler moduli 
space.
Each toric diagram represents topology of the resolution of 
${\bf C}^3/{\bf Z}_2 \times {\bf Z}_2$ in each phase.}
\label{fig:phase}
\end{center}
\end{figure}

Now we come back to the analysis of the six D-flatness 
conditions from (\ref{eq:D-flat1}) to (\ref{eq:D-flat6}).
If we consider the region $\tilde \theta_2<0$ instead of 
$\tilde \theta_2>0$, we can eliminate $p_5$ instead of $p_8$ by 
using the D-flatness condition (\ref{eq:D-flat2}).
In this case, we obtain the following three D-flatness 
conditions.
\begin{eqnarray}
|p_1|^2+|p_4|^2-|p_8|^2-|p_6|^2 &=&
-\tilde \theta_1+(-\tilde \theta_2)+\tilde \theta_3,
\label{eq:D-flat7}\\
|p_2|^2-|p_4|^2+|p_8|^2-|p_6|^2 &=&
\tilde \theta_1-(-\tilde \theta_2)+\tilde \theta_3,
\label{eq:D-flat8}\\
|p_3|^2-|p_4|^2-|p_8|^2+|p_6|^2 &=&
\tilde \theta_1+(-\tilde \theta_2)-\tilde \theta_3.
\label{eq:D-flat9}
\end{eqnarray}
One can see that the equations (\ref{eq:D-flat7}), 
(\ref{eq:D-flat8}), (\ref{eq:D-flat9}) have the same structure 
as the equations (\ref{eq:D-flat4}), (\ref{eq:D-flat5}), 
(\ref{eq:D-flat6}) if we take into account that we are 
considering the region $\tilde \theta_2<0$ instead of 
$\tilde \theta_2>0$.
Therefore, phase structure in this region is the same as that 
in the region $\tilde \theta_1>0$, $\tilde \theta_2>0$ and 
$\tilde \theta_3>0$.
In this way, we obtain phase structure near the orbifold point 
in the K${\rm {\ddot a}}$hler moduli space as shown in 
Figure \ref{fig:phase}.
With eight ($=2^3$) ways of eliminating the coordinates 
mentioned above, there are 32 phases in all.

The most striking feature of the result is that non-geometric 
phases do not appear in contrast to the analyses based on 
fundamental strings.
We can see that what makes non-geometric phases disappear is 
the redundancy of the homogeneous coordinates $p_a$.
At each point in the K${\rm {\ddot a}}$hler moduli space, 
we need only six coordinates out of nine to describe $\cal M$, 
but which coordinates we should choose depends on the region in 
the K${\rm {\ddot a}}$hler moduli space as stated above.
Thus it is important to clarify the physical origin of the 
redundancy of the coordinates to understand the reason that 
D-branes avoid non-geometric phases.

\section{D-branes on ${\bf C}^3/{\bf Z}^2 \times {\bf Z}^2$: 
an analysis based on $\theta$-stability}
\reseteqnum

In section 2, we have studied the vacuum moduli space of the 
worldvolume gauge theory of a D0-brane on the orbifold 
${\bf C}^3/{\bf Z}^2 \times {\bf Z}^2$ based on the method 
using toric geometry.
In this section, we re-examine the same space by using 
$\theta$-stability of quiver representations.

The reformulation arises as follows.
First, consider the space of solutions of F-flatness conditions.
Since it is invariant under the action of the complexified 
gauge group $G_{\bf C}$, one can talk about the 
$G_{\bf C}$-orbits satisfying the F-flatness conditions.
The key point is that, under certain conditions, 
the $G_{\bf C}$-orbit satisfying the F-flatness conditions 
contains a solution of D-flatness conditions.
They are $\theta$-stability conditions of quiver 
representations~\cite{King}.

To study $\theta$-stability of a quiver representation we need 
to know its subrepresentations.
A representation $R'$ is a subrepresentation of $R$ if there is 
an injective homomorphism from $R'$ to $R$.
Here a homomorphism between two representations 
$R=\{V_i,X_{ij}\}$ and $R'=\{V'_i,X'_{ij}\}$ is a collection of 
linear maps $\phi_i:V'_i \rightarrow V_i$ which satisfy 
$\phi_i X_{ij} =X'_{ij} \phi_j$ as matrices.
A quiver representation with a dimension vector $n$ is 
$\theta$-semistable if it satisfies
\begin{equation}
\sum n_i \theta_i=0,
\label{eq:stability1}
\end{equation}
and every subrepresentation with a dimension vector $n'$ 
satisfies
\begin{equation}
\sum n'_i \theta_i \geq 0.
\label{eq:stability2}
\end{equation}
Furthermore, a representation is $\theta$-stable if every 
non-trivial subrepresentations satisfies 
$\sum n'_i \theta_i > 0$.
Note that the equation (\ref{eq:stability1}) follows by taking 
traces of the D-flatness conditions (\ref{eq:D-flatness}), 
so the parameters $\theta_i$ in the expression coincide with 
the parameters appearing in the D-flatness conditions.

To find out $\theta$-stable representations, it is useful to 
consider Schur representations.
A representation $R$ is Schur if it satisfies End$R={\bf C}$.
This condition implies that Schur representations correspond to 
D-brane bound states~\cite{DFR2}.
Since $\theta$-stable representations are Schur, we obtain 
$\theta$-stable representations by considering Schur 
representations which satisfy F-flatness conditions, 
and then imposing $\theta$-stability conditions on the Schur 
representations.

In the case of ${\bf C}^3/{\bf Z}_2 \times {\bf Z}_2$, 
endomorphism of a representation with a dimension vector 
$(1,1,1,1)$ consists of four linear maps $\lambda_i : 
V_i \rightarrow V_i$.
The condition for the representation to be Schur is that the 
equations $\lambda_i X_{ij} = X_{ij} \lambda_j$ for $i \neq j$ 
have a unique solution 
$\lambda_1=\lambda_2=\lambda_3=\lambda_4=\lambda \in {\bf C}$.
It requires that all the nodes in the quiver diagram must be 
connected by arrows corresponding to non-vanishing $X_{ij}$'s.
Under this condition, we fix three $X_{ij}$'s to be zero by 
using the complexified gauge symmetry $G_{\rm C}=GL(3,{\bf C})$.
Since we also impose F-flatness conditions 
$X_{ik}X_{kj}=X_{il}X_{lj}$ with $k \neq l$, more than three 
$X_{ij}$'s vanish in general.
The least number of non-vanishing $X_{ij}$'s is three, 
which corresponds to the dimension of the orbifold.
To be consistent with the F-flatness conditions,
any neighboring two arrows out of the three
are in opposite directions.
By examining the conditions in detail, we found that there 
are 32 Schur representations with three non-vanishing 
$X_{ij}$'s.
We will illustrate these representations by eliminating arrows 
corresponding to vanishing $X_{ij}$'s from the quiver diagram.
The 32 Schur representations are classified into three types 
according to their forms as graphs.
\begin{figure}[htbp]
\begin{center}
\begin{picture}(400,120)
\put(43,3){(a)}
\put(143,3){(b)}
\put(243,3){(c)}
\put(343,3){(d)}
\put(44,96){\large $v_0$}
\put(10,60){\large $v_1$}
\put(44,30){\large $v_2$}
\put(80,60){\large $v_3$}
\put(144,96){\large $v_0$}
\put(110,60){\large $v_1$}
\put(144,30){\large $v_2$}
\put(180,60){\large $v_3$}
\put(244,96){\large $v_0$}
\put(210,60){\large $v_1$}
\put(244,30){\large $v_2$}
\put(280,60){\large $v_3$}
\put(344,96){\large $v_0$}
\put(310,60){\large $v_1$}
\put(344,30){\large $v_2$}
\put(380,60){\large $v_3$}
\thicklines
\put(50,90){\vector(0,-1){50}}
\put(43,93){\vector(-1,-1){24}}
\put(57,93){\vector(1,-1){24}}
\put(150,40){\vector(0,1){50}}
\put(118,68){\vector(1,1){24}}
\put(182,68){\vector(-1,1){24}}
\put(250,90){\vector(0,-1){50}}
\put(257,93){\vector(1,-1){24}}
\put(218,57){\vector(1,-1){24}}
\put(350,90){\vector(0,-1){50}}
\put(325,64){\vector(1,0){50}}
\put(357,93){\vector(1,-1){24}}
\end{picture}
\caption{Schur representations with a dimension vector 
$(1,1,1,1)$.
Arrows from $v_i$ to $v_j$ represent non-vanishing $X_{ij}$'s.
(a) The first type of Schur representations. Three arrows start 
from one node.
(b) The second type of Schur representations. Three arrows end 
at one node.
(c)(d) The third type of Schur representations.
Three arrows connecting four nodes form a line.}
\label{fig:schur}
\end{center}
\end{figure}
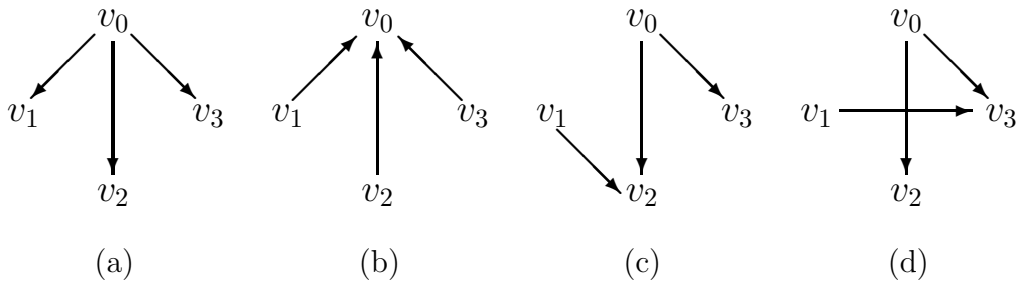
The first type of Schur representations are such that three 
arrows start from one node as shown in 
Figure \ref{fig:schur}(a).
By exchanging labels of nodes, we obtain four Schur 
representations of this type.
The Schur representation in Figure \ref{fig:schur}(a) 
has seven subrepresentations with the following dimension 
vectors,
\begin{equation}
(0,1,1,1), (0,0,1,1), (0,1,0,1), (0,1,1,0), (0,1,0,0), 
(0,0,1,0), (0,0,0,1),
\label{eq:subrepresentationA}
\end{equation}
which are determined by the condition that
an injective homomorphism exists: when $X_{ij}$ 
is non-vanishing, $n_j=0$ implies $n_i=0$. 
Hence $\theta$-stability of the representation requires
\begin{eqnarray}
\theta_1+\theta_2+\theta_3>0,~
\theta_2+\theta_3>0,~ 
\theta_3+\theta_1>0, \nonumber\\
\theta_1+\theta_2>0,~
\theta_1>0,~
\theta_2>0,~
\theta_3>0.
\end{eqnarray}
Thus the representation in Figure \ref{fig:schur}(a) is 
$\theta$-stable in the region 
$\theta_1>0,~\theta_2>0,~\theta_3>0$.
They are rewritten as 
$\tilde \theta_2+\tilde \theta_3>0,~
\tilde \theta_3+\tilde \theta_1>0,~
\tilde \theta_1+\tilde \theta_2>0$.

The second type of Schur representations are such that three 
arrows end at one node as shown in Figure \ref{fig:schur}(b).
There are four Schur representations of this type.
The representation in Figure \ref{fig:schur}(b) has 
seven subrepresentations,
\begin{equation}
(1,0,0,0), (1,1,0,0), (1,0,1,0), (1,0,0,1), (1,0,1,1), 
(1,1,0,1), (1,1,1,0),
\label{eq:subrepresentationB}
\end{equation}
and hence $\theta$-stability of the representation requires
\begin{eqnarray}
\theta_1+\theta_2+\theta_3<0,~
\theta_1+\theta_2<0,~
\theta_1+\theta_3<0, \nonumber\\
\theta_2+\theta_3<0,~ 
\theta_1<0,~
\theta_2<0,~
\theta_3<0,
\end{eqnarray}
where we have used $\theta_0+\theta_1+\theta_2+\theta_3=0$.
Thus the representation in Figure \ref{fig:schur}(b) is 
$\theta$-stable in the region 
$\theta_1<0,~\theta_2<0,~\theta_3<0$ 
($\tilde \theta_2+\tilde \theta_3<0,~
\tilde \theta_3+\tilde \theta_1<0,~
\tilde \theta_1+\tilde \theta_2<0$).

The third type of Schur representations are such that three 
arrows connecting four nodes form a line as shown in 
Figure \ref{fig:schur}(c) and \ref{fig:schur}(d).
Permutation of labels of nodes gives 24 representations of this 
type.
The representation in Figure \ref{fig:schur}(c) has six 
subrepresentations,
\begin{equation}
(0,1,1,1), (0,0,1,1), (1,0,1,1), (0,0,1,0), 
(0,0,0,1), (0,1,1,0),
\label{eq:subrepresentationC1}
\end{equation}
and hence $\theta$-stability of the representation requires
\begin{eqnarray}
\theta_1+\theta_2+\theta_3>0,~
\theta_2+\theta_3>0, \nonumber\\
\theta_1<0,~
\theta_2>0,~ 
\theta_3>0,~
\theta_1+\theta_2>0.
\end{eqnarray}
Thus the representation is $\theta$-stable in the region 
$\theta_1+\theta_2>0,~\theta_1<0,~\theta_3>0$ 
($\tilde \theta_3>0,~
-\tilde \theta_1+\tilde \theta_2+\tilde \theta_3<0,~
\tilde \theta_1+\tilde \theta_2-\tilde \theta_3>0$).
The Schur representation in Figure \ref{fig:schur}(d) also has 
six subrepresentations
\begin{equation}
(0,1,1,1), (0,0,1,1), (1,0,1,1), (0,0,1,0), 
(0,0,0,1), (0,1,0,1).
\label{eq:subrepresentationC2}
\end{equation}
As one can see, the first five subrepresentations of 
(\ref{eq:subrepresentationC2}) coincide with those of 
(\ref{eq:subrepresentationC1}); the only differences between 
(\ref{eq:subrepresentationC1}) and 
(\ref{eq:subrepresentationC2}) are the last ones, 
$(0,1,1,0)$ and $(0,1,0,1)$.
The representation in Figure \ref{fig:schur}(d) is 
$\theta$-stable in the region 
$\theta_1+\theta_3>0,~\theta_1<0,~\theta_2>0$ 
($\tilde \theta_2>0,~
-\tilde \theta_1+\tilde \theta_2+\tilde \theta_3<0,~
\tilde \theta_1-\tilde \theta_2+\tilde \theta_3>0$).

In this way, we can study regions in which Schur 
representations are $\theta$-stable.
We found that the $\theta$-stable regions for the 32 Schur 
representations cover the $\tilde \theta$-space except for 
codimension-one marginal stability loci.
It was also found that a $\theta$-stable region for a Schur 
representation of the first or second type does not overlap 
with a $\theta$-stable region for another Schur representation.
Thus in the region 
$\tilde \theta_2+\tilde \theta_3>0,~
\tilde \theta_3+\tilde \theta_1>0,~
\tilde \theta_1+\tilde \theta_2>0$, for example, the 
$\theta$-stable representation is the one in 
Figure \ref{fig:schur}(a), and there are seven 
subrepresentations (\ref{eq:subrepresentationA}).

In contrast, $\theta$-stable regions for Schur representations 
of the third type overlap with each other.
For example, the two representations in 
Figures \ref{fig:schur}(c) and \ref{fig:schur}(d) have a 
common $\theta$-stable region, 
$\tilde \theta_2>0,~\tilde \theta_3>0$, 
$-\tilde \theta_1+\tilde \theta_2+\tilde \theta_3<0$.
Therefore, a set of subrepresentations in this region
is the union of (\ref{eq:subrepresentationC1}) and 
(\ref{eq:subrepresentationC2}):
\begin{equation}
(0,1,1,1), (0,0,1,1), (1,0,1,1), (0,0,1,0), 
(0,0,0,1), (0,1,1,0), (0,1,0,1).
\label{eq:subrepresentationC}
\end{equation}

\begin{figure}[htdp]
\begin{center}
\leavevmode
\epsfysize=50mm
\epsfbox{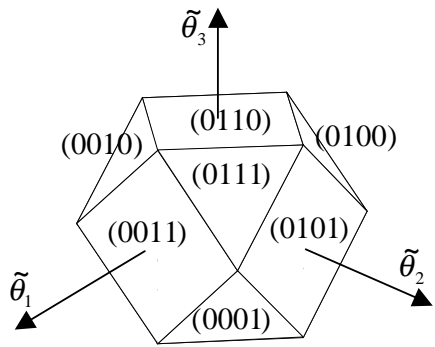}
\caption{A cuboctahedron useful to read off the structure of 
subrepresentations. 
We assign fourteen non-trivial subrepresentations of 
$(1,1,1,1)$ to fourteen surfaces. 
Sum of the dimension vectors on surfaces facing
each other is $(1,1,1,1)$.}
\label{fig:cubocta}
\end{center}
\end{figure}

In this way, we can investigate the structure of 
subrepresentations for any point in the $\tilde \theta$-space.
We found that there are 32 phases in all, just as in section 2.
In fact, there is a simple method to read off the structure of 
subrepresentations.
First consider a cuboctahedron, a polyhedron with six squares 
and eight triangles obtained from a cube by truncating its 
eight apices.
Next put it at the origin of the $\tilde \theta$-space in such 
a way that squares face to three axes 
$\tilde \theta_1$, $\tilde \theta_2$ and $\tilde \theta_3$.
Then assign fourteen non-trivial subrepresentations of 
$(1,1,1,1)$ to fourteen surfaces of the cuboctahedron in the 
following way.
Let us define 
$\tilde n_1=-n_1+n_2+n_3,~
\tilde n_2=n_1-n_2+n_3,~
\tilde n_3=n_1+n_2-n_3$.
A subrepresentation $(0,n_1,n_2,n_3)$ is assigned to the surface
with a normal vector $(\tilde n_1,\tilde n_2,\tilde n_3)$ and 
a subrepresentation $(1,1-n_1,1-n_2,1-n_3)$ to the surface 
with a normal vector $(-\tilde n_1,-\tilde n_2,-\tilde n_3)$.
The assignment is drawn in Figure \ref{fig:cubocta}.

If we assume the size of the cuboctahedron to be 
infinitesimally small, seven surfaces of the cuboctahedron are 
visible from a point in the $\tilde \theta$-space in general.
Subrepresentations of a $\theta$-stable representation 
$(1,1,1,1)$ at this point are those assigned to surfaces 
visible from the point.
A set of visible surfaces, and hence that of subrepresentations,
changes at some loci in the $\tilde \theta$-space.
Let us investigate such transitions in detail.

\subsection{Transitions corresponding to flops}

\begin{figure}[htdp]
\begin{center}
\leavevmode
\epsfysize=50mm
\epsfbox{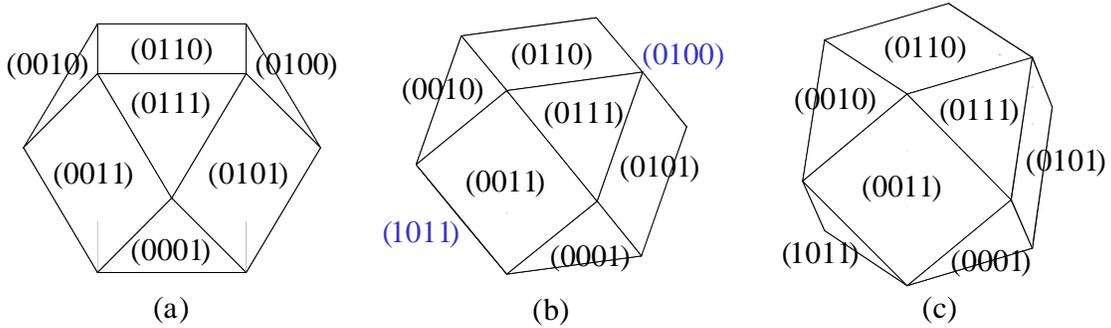}
\caption{Appearance of the cuboctahedron when one is 
(a) in the region 
$-\tilde \theta_1+\tilde \theta_2+\tilde \theta_3>0,~
\tilde \theta_1-\tilde \theta_2+\tilde \theta_3>0,~
\tilde \theta_1+\tilde \theta_2-\tilde \theta_3>0$, 
(b) on the locus 
$-\tilde \theta_1+\tilde \theta_2+\tilde \theta_3=0,~
\tilde \theta_2>0,~\tilde \theta_3>0$, and 
(c) in the region 
$-\tilde \theta_1+\tilde \theta_2+\tilde \theta_3<0,~
\tilde \theta_2>0,~\tilde \theta_3>0$.}
\label{fig:flop}
\end{center}
\end{figure}

First we would like to examine a process corresponding to a 
flop transition.
Consider a point in the region 
$-\tilde \theta_1+\tilde \theta_2+\tilde \theta_3>0,~
\tilde \theta_1-\tilde \theta_2+\tilde \theta_3>0,~
\tilde \theta_1+\tilde \theta_2-\tilde \theta_3>0$, 
from which seven surfaces corresponding to 
(\ref{eq:subrepresentationA}) are visible as shown in 
Figure \ref{fig:flop}(a).
As the value of $\tilde \theta_1$ increases, the area of the 
surface $(0,1,0,0)$ visible from that point becomes smaller, 
and it degenerate to a line when one comes to the locus 
$-\tilde \theta_1+\tilde \theta_2+\tilde \theta_3=0$.
At the same time the surface $(1,0,1,1)$ opposite to 
$(0,1,0,0)$ becomes visible as a line as in 
Figure \ref{fig:flop}(b).
As one moves further, the surface $(0,1,0,0)$ completely 
disappears while the surface $(1,0,1,1)$ appears as in 
Figure \ref{fig:flop}(c).
Thus $(1,0,1,1)$ replaces $(0,1,0,0)$ as a subrepresentation 
when one crosses the locus 
$-\tilde \theta_1+\tilde \theta_2+\tilde \theta_3=0$.

Compared with what occurs geometrically at the same locus,
this process should correspond to a flop transition
$[C_1]\rightarrow -[C_1]$, as in section 2.
In fact, if we consider the following correspondence between 
representations and homology classes,
\begin{eqnarray}
(0,1,0,0) \Leftrightarrow [C_1],\nonumber \\
(0,0,1,0) \Leftrightarrow [C_2],
\label{eq:Mckay}\\
(0,0,0,1) \Leftrightarrow [C_3],\nonumber
\end{eqnarray}
the process exchanging $(0,1,0,0)$ for $(1,0,1,1)$ is 
interpreted as a flop.
The key point is the fact that $(1,1,1,1)$ corresponds to a 
D0-brane as noted in section 2.
Thus the representation $(1,0,1,1)=(1,1,1,1)-(0,1,0,0)$ should 
correspond to a homology class $[p]-[C_1]$ where $[p]$ 
represents a homology class of a point corresponding to a 
D0-brane.
If we ignore $[p]$ \footnote{To exactly discuss the process, 
we should use K-theory instead of homology theory.
In K-theory terms, the flop transition corresponds to an 
exchange $[k_1] \leftrightarrow -[k_1]$, where $[k_i]$ 
represents a generator of K-theory~\cite{AP}.}, 
the exchange $(0,1,0,0) \leftrightarrow (1,0,1,1)$ implies 
$[C_1] \leftrightarrow -[C_1]$, which is nothing but the flop 
transition discussed in section 2.
The correspondence (\ref{eq:Mckay}) agrees 
with the result obtained from the other methods~\cite{AP,OFS}. 

Here we comment on an implication of the result on the McKay 
correspondence.
The McKay correspondence between representations of $\Gamma$ 
and geometry of resolutions of ${\bf C}^3/\Gamma$ 
has been mainly discussed for a particular resolution of the orbifold 
called a Hilbert scheme.
In the present case, the Hilbert scheme is the space with a 
topology represented by the toric diagram in 
Figure \ref{fig:toric}(a).
The above result can be interpreted as providing a way to calculate 
the McKay correspondence explicitly for resolutions other than the 
Hilbert scheme\footnote{After we completed this work, 
we were informed that the McKay correspondence for 
resolutions other than the Hilbert scheme was 
explicitly investigated in the cases 
$\Gamma={\bf Z}_2 \times {\bf Z}_2, {\bf Z}_6, {\bf Z}_{11}$ 
in \cite{Craw}.}.

\subsection{Transitions corresponding to change of coordinates}

\begin{figure}[htdp]
\begin{center}
\leavevmode
\epsfysize=50mm
\epsfbox{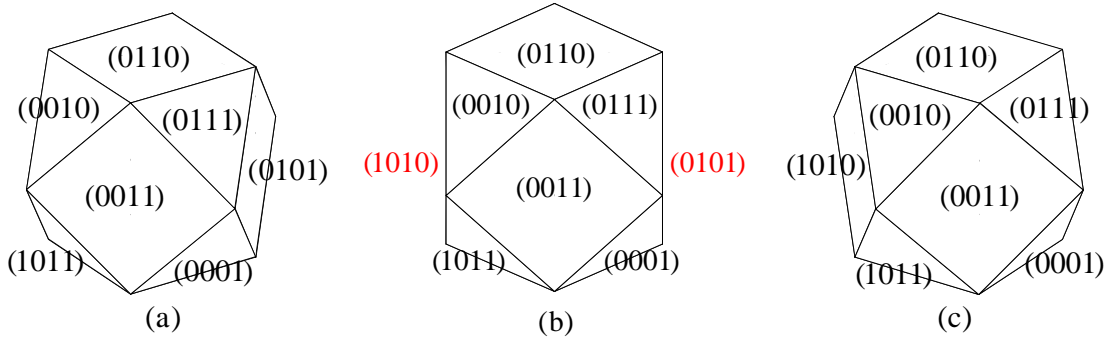}
\caption{Appearance of the cuboctahedron 
(a) in the region 
$\tilde \theta_2>0,~\tilde \theta_3>0,~
\tilde \theta_1-\tilde \theta_2-\tilde \theta_3>0$, 
(b) on the locus 
$\tilde \theta_2=0,~\tilde \theta_3>0,~
\tilde \theta_1-\tilde \theta_3>0$, and 
(c) in the region 
$\tilde \theta_2<0,~\tilde \theta_3>0,~
\tilde \theta_1+\tilde \theta_2-\tilde \theta_3>0$.}
\label{fig:monodromy}
\end{center}
\end{figure}

Next we would like to investigate another type of transition.
Start with a point in the region 
$\tilde \theta_2>0,~\tilde \theta_3>0$ and 
$-\tilde \theta_1+\tilde \theta_2+\tilde \theta_3<0$, 
from which seven surfaces corresponding to 
(\ref{eq:subrepresentationC}) are visible as shown in 
Figure \ref{fig:monodromy}(a).
As the value of $\tilde \theta_2$ decreases, the area of the 
surface $(0,1,0,1)$ visible from that point becomes smaller, 
and it degenerate to a line when one comes to the locus 
$\tilde \theta_2=0$.
At the same time the surface $(1,0,1,0)$ opposite to 
$(0,1,0,1)$ becomes visible as a line as in 
Figure \ref{fig:monodromy}(b).
As one moves further, the surface $(0,1,0,1)$ completely 
disappears while the surface $(1,0,1,0)$ appears as in 
Figure \ref{fig:monodromy}(c).
Thus $(1,0,1,0)$ replaces $(0,1,0,1)$ as a subrepresentation 
when one crosses the locus $\tilde \theta_2=0$.

Compared with the result in section 2, this process should 
correspond to a change of homogeneous coordinates from 
$p_5$ to $p_8$.
For a consistent interpretation of the process, it is found to 
be necessary to consider the following permutation of 
subrepresentations,
\begin{equation}
(0,0,1,0) \leftrightarrow (0,1,1,1), \quad 
(0,0,0,1) \leftrightarrow (1,0,1,1),
\end{equation}
in addition to the exchange $(0,1,0,1)$ for $(1,0,1,0)$.
This transformation of subrepresentations is nothing but the 
reflection of the cuboctahedron with respect to the plane 
$\tilde \theta_2=0$.
Similarly, the exchange of the homogeneous coordinates 
$p_4 \leftrightarrow p_7$ ($p_6 \leftrightarrow p_9$) 
in the toric method corresponds to the reflection with respect 
to $\tilde \theta_1=0$ ($\tilde \theta_3=0$).
We denote the reflection with respect to the plane 
$\tilde \theta_a=0$ as $A_a$.

To understand the meaning of the exchange of the homogeneous 
coordinates, we consider the combinations of the reflections; 
$M_1=A_2 A_3,~M_2=A_3 A_1,~M_3=A_1 A_2$.
For example, $M_1$ corresponds to the action 
$\tilde \theta_2 \rightarrow -\tilde \theta_2,~
\tilde \theta_3 \rightarrow -\tilde \theta_3$, 
and leads to the following permutation of subrepresentations,
\begin{eqnarray}
(1,0,0,0) \leftrightarrow (0,1,0,0),\nonumber\\
(0,0,1,0) \leftrightarrow (0,0,0,1),\nonumber\\
(0,1,1,1) \leftrightarrow (1,0,1,1),\nonumber\\
(1,1,0,1) \leftrightarrow (1,1,1,0),\\
(1,0,1,0) \leftrightarrow (0,1,0,1),\nonumber\\
(1,0,0,1) \leftrightarrow (0,1,1,0).\nonumber
\end{eqnarray}

Based on the correspondence between representations and 
homology classes obtained above, the action $M_1$ is 
geometrically interpreted as
\begin{equation}
[C_1] \leftrightarrow [p]-[C_1]-[C_2]-[C_3], \quad 
[C_2] \leftrightarrow [C_3].
\end{equation} 
Similarly, the actions $M_2$ and $M_3$ are represented in 
terms of geometry.
It is known that the actions $M_i$ satisfying the relations, 
$M_1^2=M_2^2=M_3^2=1, \, M_1 M_2=M_3$, are interpreted as 
monodromies around the orbifold point in the 
K${\rm {\ddot a}}$hler moduli space~\cite{OFS}.

Thus the change of homogeneous coordinates in the toric method 
stems from the orbifold monodromy, and the number $c$ of the 
coordinates in the expression (\ref{eq:moduli}) can be 
understood from this viewpoint.
As discussed in section 2, the fact that D-branes do not see 
non-geometric phases comes from the redundancy of the 
homogeneous coordinates.
Thus the above argument shows that this property is a 
consequence of the monodromy around the orbifold point.

For a few examples other than 
${\bf C}^3/{\bf Z}_2 \times {\bf Z}_2$, we have verified that 
the redundancy of homogeneous coordinates in the expression 
(\ref{eq:moduli}) can be understood from orbifold monodromy. 
It would be interesting to have a general expression for the 
number $c$ of homogeneous coordinates in the expression 
(\ref{eq:moduli}) for orbifolds ${\bf C}^3/{\bf Z}_n$ and 
${\bf C}^3/{\bf Z}_n \times {\bf Z}_m$.

\subsection{Decay products on marginal stability loci}

Finally we would like to discuss decay products at marginal 
stability loci.
They are read off from the cuboctahedron as follows.
Subrepresentations on its surfaces visible from a point in the 
$\tilde \theta$-space are interpreted as potential decay 
products.
As one of the surfaces shrinks to a line, decay is triggered by 
the subrepresentation on this surface since the stability 
condition (\ref{eq:stability2}) for this subrepresentation 
saturates.
Remaining elements of the decay must be such that the sum of 
dimension vectors of decay products is $(1,1,1,1)$.
By inspecting the rule of assignment of subrepresentations on 
the cuboctahedron, one can see that decay products are 
two subrepresentations on the shrinking surfaces at the 
marginal stability loci.
For example, on the locus 
$-\tilde \theta_1+\tilde \theta_2+\tilde \theta_3=0$, 
$(1,1,1,1)$ decays into $(0,1,0,0)$ and $(1,0,1,1)$ as shown in 
Figure $\ref{fig:flop}$(b).

This result agrees with the conjecture on decay products given 
in \cite{FM}.
To explain the conjecture, we need to make some definitions.
A sequence of subrepresentations of a semi-stable 
representation $R$,
\begin{equation}
0=R_0 \subset R_1 \subset R_2 \subset ... \subset R_m =R,
\end{equation}
is called a Jordan-H${\rm {\ddot o}}$lder filtration if the 
dimension vectors $n_a$ of $R_a$ satisfy 
$\theta \cdot n_a=0$ and the quotients $M_a=R_a/R_{a-1}$ are 
$\theta$-stable.
By using a Jordan-H${\rm {\ddot o}}$lder filtration, 
the graded representation of $R$ is defined as
\begin{equation}
{\rm gr}(R)=\oplus_a M_a.
\label{eq:graded}
\end{equation}
Given a semi-stable representation $R$, there may be several 
Jordan-H${\rm {\ddot o}}$lder filtrations but ${\rm gr}(R)$ is 
unique.
We also note that ${\rm gr}(R)$ coincides with $R$ for 
$\theta$-stable representations.
The conjecture given in \cite{FM} is that decay products of $R$ 
on marginal stability loci are given by ${\rm gr}(R)$.
In the present case, on the locus 
$-\tilde \theta_1+\tilde \theta_2+\tilde \theta_3=0$, 
the representation $R=(1,1,1,1)$ has a 
Jordan-H${\rm {\ddot o}}$lder filtration,
\begin{equation}
(0,0,0,0) \subset (0,1,0,0) \subset (1,1,1,1).
\end{equation}
It leads ${\rm gr}(R)=(0,1,0,0) \oplus (1,0,1,1)$.
Thus D0-brane decays into a threshold bound state of 
$(0,1,0,0)$ and $(1,0,1,1)$ at 
$-\tilde \theta_1+\tilde \theta_2+\tilde \theta_3=0$, 
in agreement with the result read off from the cuboctahedron.

\section{Discussions}
\reseteqnum

In this section, we would like to discuss why non-geometric 
phases are projected out from the viewpoint of stability of 
D0-branes\footnote{In \cite{AL}, a relation between the 
non-geometric phase and 0-brane instability was argued in a 
different context.}.

In standard classical geometry, elementary constituents of a
space are points; correspondingly, a space probed by a point 
particle is described by classical geometry.
In this paper, we have used a D0-brane as a probe to study the 
orbifold.
Since a D0-brane is a point-like object, it may seem reasonable 
that a space probed by a D0-brane has a classical geometric 
interpretation.
However, there is an issue we have to care about: stability of 
a D0-brane.
If a D0-brane becomes unstable, it decays into a sum of higher 
dimensional objects in general, and loses the property as a 
point-like probe.
Thus the most likely expectation is that a space probed by a 
D0-brane is described by classical geometry only if the 
D0-brane is stable.
This expectation is consistent with the result in this paper.
As we have seen, any point in the $\tilde \theta$-space where a 
D0-brane is stable belongs to geometric phases, and the loci on 
which D0-brane is marginally stable coincide with boundaries of 
geometric phases.
In this respect, the fact that D0-branes are always 
(semi) stable may be understood as one explanation of the 
disappearance of non-geometric phases.

Of course, it is not clear that the expectation holds for other 
cases.
It is necessary to examine to what extent the expectation can 
be applied.

\vskip 1cm
\centerline{\large\bf Acknowledgements}

We would like to thank Y. Ito, Y. Ohtake and T. Takayanagi for 
valuable discussions.

\end{document}